%% file: main.tex
\documentclass[sigconf]{acmart}

\usepackage{multirow}

\usepackage{amsmath}
\usepackage{amsfonts}
\usepackage{scalerel}
\usepackage{mathtools}
\usepackage{algorithm}
\usepackage{algorithmicx}

\usepackage{subfigure}

\usepackage[noend]{algpseudocode}

\usepackage{enumitem}
\usepackage{booktabs}
\usepackage{adjustbox}
\usepackage{graphicx}
\usepackage[justification=justified, skip=0pt]{caption}

\usepackage{soul}
\setstcolor{red}

\makeatletter
\g@addto@macro\normalsize{%
  \abovedisplayskip 4pt plus 2pt minus 3pt%
  \belowdisplayskip \abovedisplayskip
  \abovedisplayshortskip 4pt plus 2pt minus 3pt%
  \belowdisplayshortskip 4pt plus 2pt minus 3pt%
}

\newcommand\blfootnote[1]{%
  \begingroup
  \renewcommand\thefootnote{}\footnote{#1}%
  \addtocounter{footnote}{-1}%
  \endgroup
}

\makeatother

\AtBeginDocument{%
  \providecommand\BibTeX{{%
    \normalfont B\kern-0.5em{\scshape i\kern-0.25em b}\kern-0.8em\TeX}}}

\setcopyright{acmcopyright}
\copyrightyear{2021}
\acmYear{2021}
\acmDOI{10.1145/1122445.1122456}

\acmConference[CIKM '21]{CIKM '21: 30th ACM International Conference on Information and Knowledge Management}{1--5 November 2021}{ Queensland, Australia}
\acmBooktitle{CIKM '21: ACM International Conference on Information and Knowledge Management,November 1--5, 2021 Queensland, Australia}
\acmPrice{15.00}
\acmISBN{978-1-4503-XXXX-X/18/06}

\newcommand{\bigCI}{\mathrel{\text{\scalebox{1.07}{$\perp\mkern-10mu\perp$}}}}

\newtheorem{assumption}{Assumption}
\newtheorem{definition}{Definition}

\begin{document}

\title{CausCF: Causal Collaborative Filtering for Recommendation Effect Estimation}
\author{Xu Xie$^{1,2*}$,
        Zhaoyang Liu$^{1*}$,
        Shiwen Wu$^{1,2}$,
        Fei Sun$^1$,
        Cihang Liu$^1$,
        Jiawei Chen$^{1,3}$,\\
        Jinyang Gao$^1$,
        Bin Cui$^1$,
        Bolin Ding$^2$,
        }
\affiliation{\institution{$^1$Alibaba Group,
                          $^2$Peking University,
                          $^3$University of Science and Technology of China,
                          }
            }
\email{{jingmu.lzy, dustin.lch, haoyi.yz, jinyang.gjy, bolin.ding}@alibaba-inc.com
}
\email{{xu.xie, wushw.18, bin.cui}@pku.edu.cn,  cjwustc@ustc.edu.cn 
}

\renewcommand{\shortauthors}{Xie and Liu, et al.}
\renewcommand{\authors}{Xu Xie, Zhaoyang Liu, Shiwen Wu, Cihang Liu, Fei Sun, Jiawei Chen, Jinyang Gao, Bin Cui, and Bolin Ding}

\input{subfile/0_abstract.tex}



\maketitle

\input{subfile/1_introduction.tex}
\input{subfile/2_related.tex}

\input{subfile/3_formulation.tex}
\input{subfile/4_method.tex}

\input{subfile/5_evaluation.tex}

\input{subfile/6_experiment.tex}
\input{subfile/7_conclusion.tex}



\balance
\bibliographystyle{ACM-Reference-Format}
\bibliography{sample-base}

\appendix

\end{document}

%% file: subfile/0_abstract.tex
\begin{abstract}
    
    To improve user experience and profits of corporations, modern industrial recommender systems usually aim to select the items that are most likely to be interacted with (e.g., clicks and purchases).
    However, they overlook the fact that users may purchase the items even without recommendations. 
    The real effective items are the ones which can contribute to purchase probability uplift.
    To select these effective items, it is essential to estimate the causal effect of recommendations.
    Nevertheless, it is difficult to obtain the real causal effect since we can only recommend or not recommend an item to a user at one time.
    Furthermore, previous works usually rely on the randomized controlled trial~(RCT) experiment to evaluate their performance.
    However, it is usually not practicable in the recommendation scenario due to its unavailable time consuming.
    To tackle these problems, in this paper, we propose a causal collaborative filtering~(CausCF) method inspired by the widely adopted collaborative filtering~(CF) technique. 
    It is based on the idea that similar users not only have a similar taste on items, but also have similar treatment effect under recommendations.
    CausCF extends the classical matrix factorization to the tensor factorization with three dimensions---user, item, and treatment.
    Furthermore, we also employs regression discontinuity design (RDD) to evaluate the precision of the estimated causal effects from different models.
    With the testable assumptions, RDD analysis can provide an unbiased causal conclusion without RCT experiments.
    Through dedicated experiments on both the public datasets and the industrial application,
    we demonstrate the effectiveness of our proposed CausCF on the causal effect estimation and ranking performance improvement.
\end{abstract}

%% file: subfile/1_introduction.tex
\section{Introduction}

\begin{table}[t]
    \centering
    \small
    \caption{The possibility of purchase w/o recommendations on different items.}
    \label{table:introduction_case}
    \begin{adjustbox}{max width=0.8\textwidth}
    \begin{tabular}{l|cc}
        \toprule
         & Shampoo &  Nintendo Switch \\
        \midrule
        Recommended  & 95\% & 50\%\\
        Not recommended  & 90\%  & 10\% \\
        \midrule
        Recommendation effect & 5\% &40\% \\
        \bottomrule
    \end{tabular}
    \end{adjustbox}
\end{table}

Recommender systems have achieved great success in many industrial applications, e.g., Youtube, Amazon, and Netflix.
\blfootnote{$^{*}$Equal Contribution.}
To improve user experience and profits of corporations, both industry and academia have long been pursuing better recommendation models
which can recommend items with a higher probability of click or purchase~\cite{ncf,sasrec,bert4rec}.
However, they ignore that users would still purchase the items without recommendations, especially when they have clear purchase intentions.
Taking the E-commerce platform as an example, 
it generally has a series of scenarios, including the recommender system, organic search, 
and homepages of brand sellers.
Recommending the items that the user will purchase in other scenarios would not increase the platform's positive interactions or commercial values.

To tackle this problem, we argue that instead of recommending items that customers will interact with, recommender systems need to recommend items that result in higher purchase probability lift.
To illustrate the importance of uplift, we give a simple example in e-commerce.
As shown in Table~\ref{table:introduction_case}, there exist two types of items, including shampoo and Nintendo Switch.
Traditional recommender systems prefer to suggest the shampoo for users since the shampoo has a higher purchase probability when recommended.
However, if we can also observe the possibility in the counterfactual world, recommending Nintendo Switch is a better choice since it has a higher lift of purchase probability (40\% vs. 5\%).
Therefore, the causal effects are important for enhancing the performance of recommendations, but it is difficult to estimate them.
On the one hand, there is no ground truth to estimate such improvements since we can only recommend or not recommend an item to a user at one time.
On the other hand, utilizing the statistics from the observations directly may bias the conclusion or even cause Simpson's paradox~\cite{simpson}.


To handle these challenges, we employ the causal inference framework to consider the results in both factual and counterfactual worlds. 
Causal inference has become an appealing research area and achieved promising results in many domains~\cite{survey1,survey2}.
In the context of the recommender system, 
most previous works have attempted to eliminate biases (e.g., position bias) in the recommender systems from a causal perspective~\cite{pop_bias,position_bias,rec_causal_icml,rec_causal2016}.
But the causal effect of recommendations under multiple scenarios in e-commerce platforms is less studied.
A close line with our work is uplift-based optimization which optimizes the ranking list via the uplift metrics~\cite{rec_impact,rec_uplift}.
However, these works still have some limitations.
1) the previous models fails to utilize the collaborative information hidden in the user-item interactions;
2) they generally do not distinguish the difference of causal effects on users and items;
3) the evaluation heavily relies on the randomized controlled trials~(RCT) experiments, whose time-consuming is not available in e-commerce recommendations.

In this paper, we propose a novel model called causal collaborative filtering~(CausCF), which attempts to estimate the causal effect of recommendation in a collaborative way.
Typically, the collaborative filtering~(CF) assumes that similar users prefers similar items.
Our CausCF extends the idea to the causal inference context and claim that similar users may not only have similar preferences on items, 
but have similar causal effects on the same items.
Based on the famous matrix factorization model~\cite{svd++}, CausCF naturally extends the interaction matrix to a three-dimensional tensor with the dimensionality of the user, item, and treatment.
In this way, it predicts interaction probability by the pairwise inner product of users, items, and treatments.
The inner product between user and item representations illustrate the user preference of the item like the matrix factorization.
And the inner product between user and treatment representations describe the treatment effects of users.
The treatment effects of items can be estimated as users.
By this formulation, the causal effect estimation of users and items can benefit from the collaborative information of similar users or items.
Furthermore, we propose a novel evaluation method under the recommendation scenarios.
Though the golden rule to evaluate the causal inference performance is to conduct RCT experiments, the expensive experimental cost makes it infeasible in online recommender systems.
It is unacceptable to obtain the causal effect of recommendation at the cost of sacrificing user engagements.
In this paper, we also propose to utilize a typical econometric tool, regression discontinuity analysis~(RDD), to make unbiased causal conclusions.
It estimates the causal effect by comparing the observations around the discontinuities, which are the positions where the user stops browsing.

The contributions of this paper are summarized as follows:
\begin{itemize}[noitemsep,topsep=0pt]
    \item We propose a new perspective for recommender systems. 
    Instead of focusing on the performance of the recommender system itself, 
    we suggest recommending the items which can significantly increase the purchase probabilities.
    \item By abstracting the problem as a tensor factorization task, 
    we propose a causal collaborative filtering method to estimate the user and item-specific causal effect on individuals. 
    \item With the position information of user browsing, 
    we propose a practical evaluation method using regression discontinuity analysis to make unbiased causal conclusions.
    \item The experiments conducted on both public datasets and industrial applications demonstrate the effectiveness of our method.
    The precision of the estimated effect and the ranking performance are improved by 13.8\% and 21.3\%, respectively, compared with the competitive baselines.
\end{itemize}

%% file: subfile/2_related.tex
\section{Related Work}

\subsection{Causal Effect Estimation}

As summarized by \citet{survey2}, 
methods related to the causal effect estimation can be divided into two major categories according to whether requiring the ignorability assumption or not.
With the ignorability assumption, 
typical methods include regression adjustment, propensity score re-weighting, and covariate balancing.
Regression adjustment~\cite{harmless_econ} generally constructs regression equations to align input features and predict outcomes under different treatments.
The propensity score is used to balance the distributions of the treated and control groups~\cite{snips,psm_intro,psm_survey}.
By re-weighting the data samples with the estimated propensity score,
we can alleviate the selection biases in observations.
As for covariate balancing, 
a series of works solve the problem by reformulating the counter-factual estimation problem as the domain adaptation task\cite{ace,yao2018representation}.
Moreover, \citet{tarnet} have proved that the estimated causal effect error can be bounded by the generalization loss and the distribution distance between the treated and control groups.

When relaxing the ignorability assumption, it is common to introduce the extra variables or utilize alternative information.
Two popular methods are instrumental variable (IV) method~\cite{instrumental_96,instrumental_14} and regression discontinuity design (RDD)~\cite{rd_2008,rd_2010}.
As the term suggests, the effectiveness of IV and RDD methods heavily relies on the choice of instrumental variables and discontinuities.
The instrumental variables are required to be the preceding variables of the treatment and affect the outcome only by the treatment assignment.
The causal effect is then estimated by constructing a two-stage regression where the first-stage regression is to measure how much a certain treatment changes if we modify the relevant IVs
and the second-stage regression is to measure the changes in the treatment caused by IV would influence the outcomes~\cite{2sls,deepiv}.
Sometimes, treatment assignments may only depend on the value of a special feature, which is generally named as the running variable~\cite{rdd_review,rdd_pitfall}.
And the causal effect can be induced by comparing observations around the value of the running variable~\cite{rdd_automated}.

\subsection{Causal Inference in Recommendation}

For recommendation, 
most existing works focus on the idea of using causal inference to eliminate biases, e.g., popularity bias, position bias, and exposure bias~\cite{rank_bias,rec_causal_icml,unbiased_psw,double_miss}.
The main reason is that the observations of user feedback are always conditioned on item exposure or displayed positions.
\citet{rank_bias} proposed estimating the selection bias through a randomization experiment and used the inversed propensity score (IPS) to debias the training loss computed with the click signals.
A doubly robust way is also proposed to integrate the imputed errors and propensities to alleviate the influence
of the propensity variance~\cite{double_miss}.
\citet{unbias_ipw} analyzed the IPS framework and showed that it could realize unbiased learning theoretically and empirically.
Compared with these works, we devote ourselves to estimating the causal effect of recommendations rather than debiasing user feedback.

Recently, recommendation strategies targeting the causal effect arise more attentions~\cite{rec_impact,causal_emb,rec_uplift}.
In general, recommender systems have a positive effect on the e-commercial ecosystem,
benefiting business values such as gross merchandise value (GMV) and user engagement\-~\cite{business_value}.
Recommendations naturally increase the probability that an item will be clicked or purchased on the platform.
\citet{rec_uplift} directly modeled such increases by formalizing the problem as a binary classification task and re-labeling the samples with the potential of increases.
\citet{causal_emb} proposed to train two separate prediction models: 
one with the recommendation and the other without, by imposing regularization on parameters.
The causal effect is then estimated by the difference between the prediction values.
These methods have achieved promising results in public datasets,
but the evaluation is trivial or heavily relies on randomized experiments.
In this paper, we attempt to reformulate the problem from the tensor perspective
and model the user and item-specific treatment effect, respectively.
Moreover, we evaluate the causal conclusions via RDD analysis.

%% file: subfile/3_formulation.tex
\section{Preliminary}

In this section, we will describe notations under the potential outcome framework, present the problem definition and introduce several common assumptions in causal inference.

\subsection{Notations and Problem Definition}

The potential outcome framework defines several key components, including the unit $n$, the treatment $T$, and the outcome $Y$. 
In the context of recommendation,
these components can be described as follows:
1) The unit $n$, as the atomic research object in the study, is defined by each pair of user and item.
2) The treatment $T$ is a binary variable $T=\{0, 1\}$, indicating the recommendation exposure or not.
3) For each unit-treatment pair, 
the potential outcome $Y(T=t)$ is the purchase probability under different recommendation treatments.
Moreover, there exist two types of information for units, e.g., pre-treatment variables and post-treatment variables, 
to help identify the causal effect more precisely.
Post-treatment variables are the variables affected by the treatment, such as post-click and dwell time.
Pre-treatment variables are those that will not, including the user profile and item descriptions.
In the following sections, we will focus on the pre-treatment variables since they are not influenced by the treatment and used for the potential outcome prediction.
We denote the unit with pre-treatment variables $x$ as $X=x$.
With the pre-treatment variables of the unit, the potential outcome can be formulated as a function of the treatment $t$ and pre-treatment variable $x$, 
that is, $Y(T=t|X=x)$.

Based on the notations mentioned above, we can define the individual treatment effect (ITE) for each pair of user and item
by comparing the treatment under $T=1$ and $T=0$,
\begin{equation}
    \mathrm{ITE}(x)=Y(T=1|X=x)-Y(T=0|X=x).
\end{equation}
By aggregating the results of ITE, we can get the causal conclusions at different population levels, e.g., 
the Conditional Average Treatment Effect (CATE) in different age group or the Average Treatment Effect (ATE) among the whole population.
In this paper, we define our problems as:

\begin{definition}[Causal Effect Estimation for Recommendation.]
    Given the pre-treatment variables $x$ for each pair of user and item,
    we try to predict the potential purchase probability $Y(T=t|X=x)$ under different recommendation treatments.
    The causal effect is then computed by comparing the potential outcomes $Y(T=1|X=x)$ and $Y(T=0|X=x)$, 
    and then aggregated at different population levels.
\end{definition}

\subsection{Causal Assumptions}\label{sec:assumption}
Since a unit can only take one treatment at one time, 
we can not obtain the counterfactual outcome from observations.
To estimate the causal effect, we rely on the following assumptions that are widely used in the causal inference.
\begin{assumption}
\textbf{Stable Unit Treatment Value Assumption (SUTVA)}. Units do not interfere with each other, and treatment levels are well defined.

It requires that the purchase probabilities for any user-item pair are independent of the treatments which other user-item pairs receive, 
and the treatment levels are the same for different units.
\end{assumption}

\begin{assumption}
    \textbf{Ignorability}. Given the pre-treatment variable $X$, treatment assignment $T$ is independent to its potential outcomes, with the mathematical formulation as:
\begin{equation}
    T \bigCI Y(T=t)  | X=x.
\end{equation}

It means that given the user profile and item descriptions, the recommendation assignment is independent of its potential purchase probability on the platform.
\end{assumption}

\begin{assumption}
    \textbf{Positivity}. For any set of values of pre-treatment variable $X$, treatment assignment is not deterministic
    \begin{equation}
        0 < P(T=t|X=x) < 1, \forall t \text{ and } x.
    \end{equation}

    Under the recommender system context, each user-item pair should be possible to be recommended.
\end{assumption}

Although these assumptions are hard to be verified, we follow these assumptions to design causal methods.

%% file: subfile/4_method.tex
\begin{figure}[t]
    \centering
    \includegraphics[width=.65\linewidth]{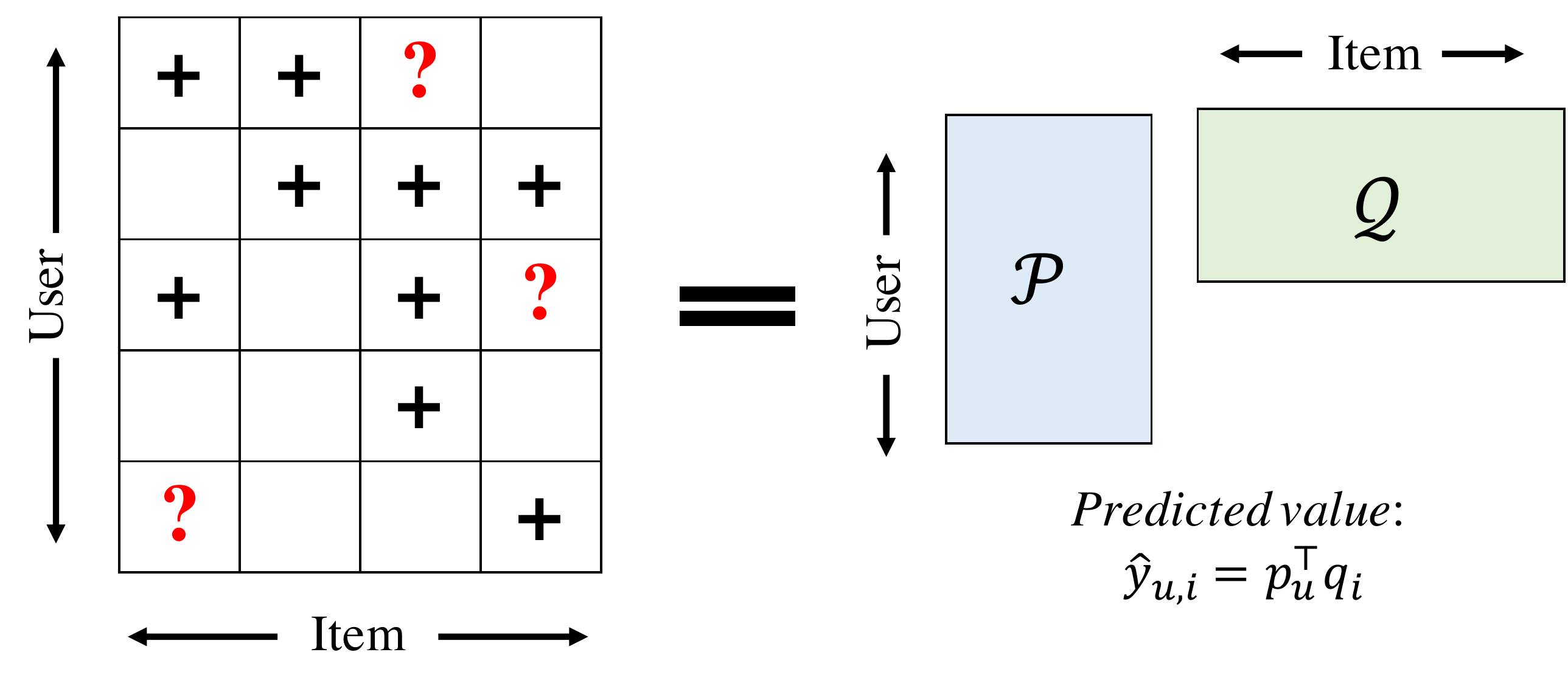}
    \caption{Matrix factorization illustration.}
    \label{fig:mf_model}
\end{figure}

\section{Methodology}

In this section, we will first review the latent factor model of collaborative filtering 
and then present our proposed causal collaborative filtering method named CausCF.

\subsection{Collaborative Filtering}

\begin{figure*}[t]
    \centering
    \subfigure[Tensor factorization for causal collaborative filtering.]{
        \includegraphics[width=0.3\linewidth]{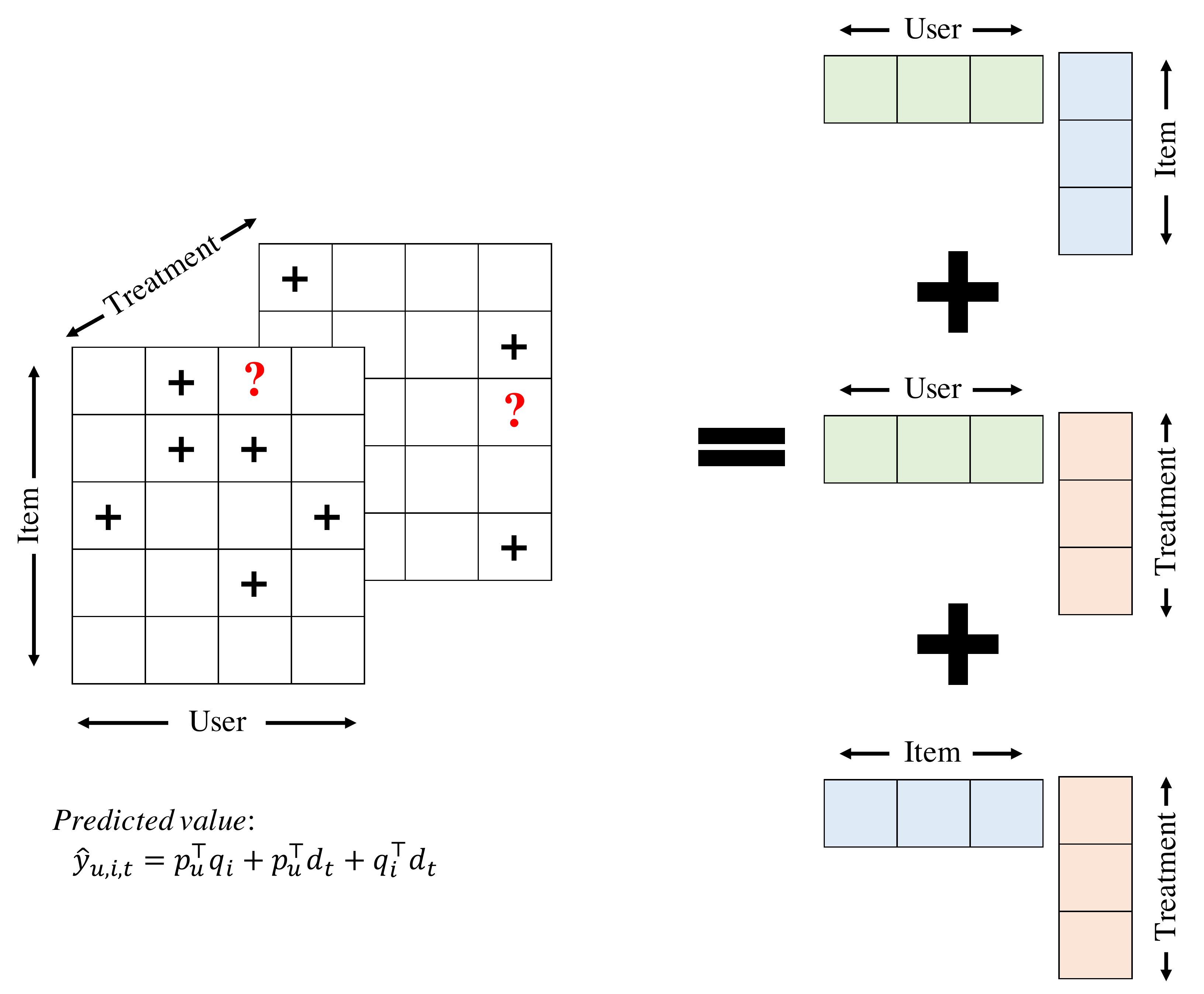}
        \label{fig:infer_model}
    }
    \hfil
    \subfigure[Model design for latent representation $p_u$.]{
        \includegraphics[width=0.25\linewidth]{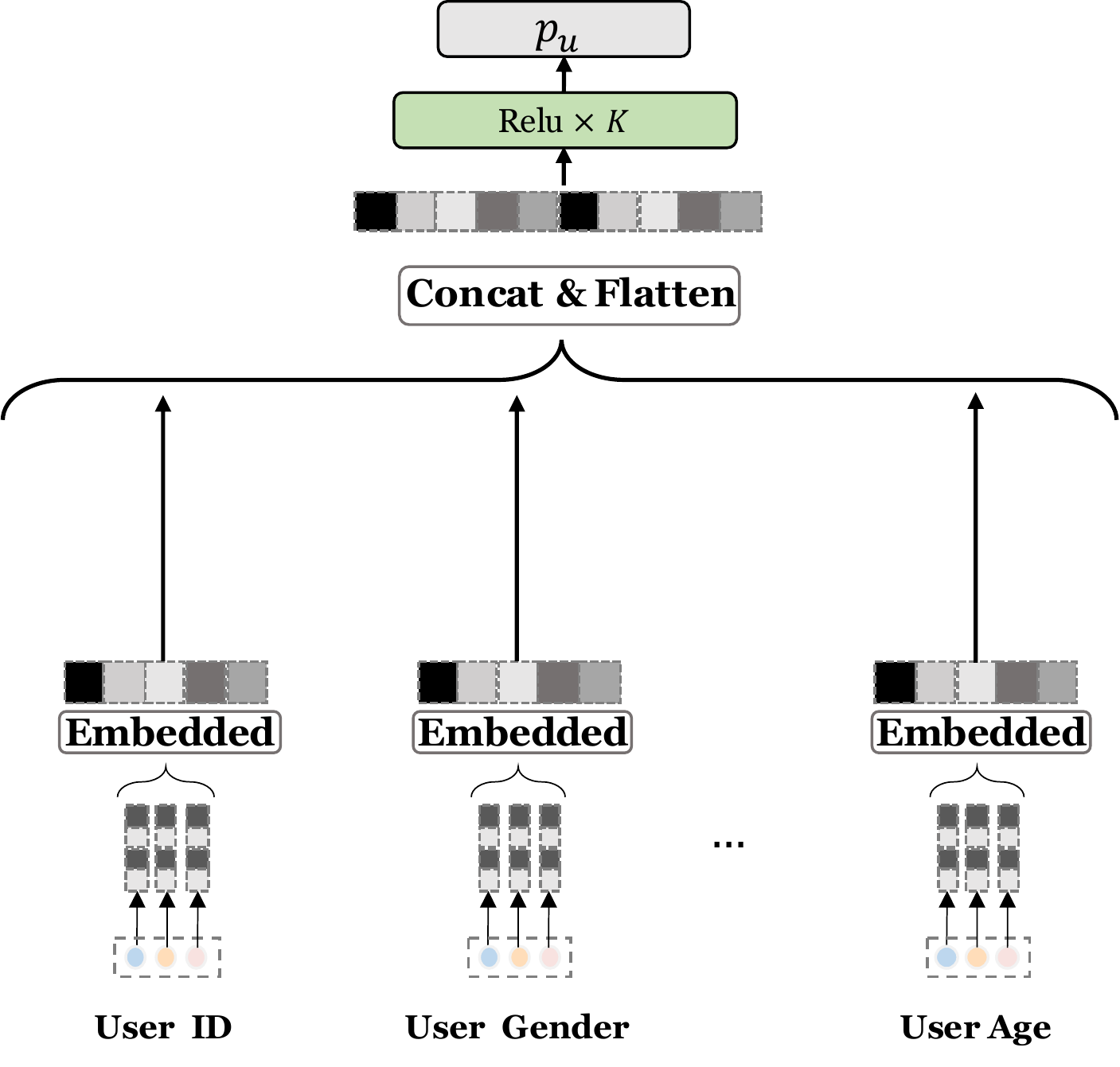}
        \label{fig:user_structure}
    }
    \hfil
    \subfigure[Model design for latent representation $q_i$.]{
        \includegraphics[width=0.25\linewidth]{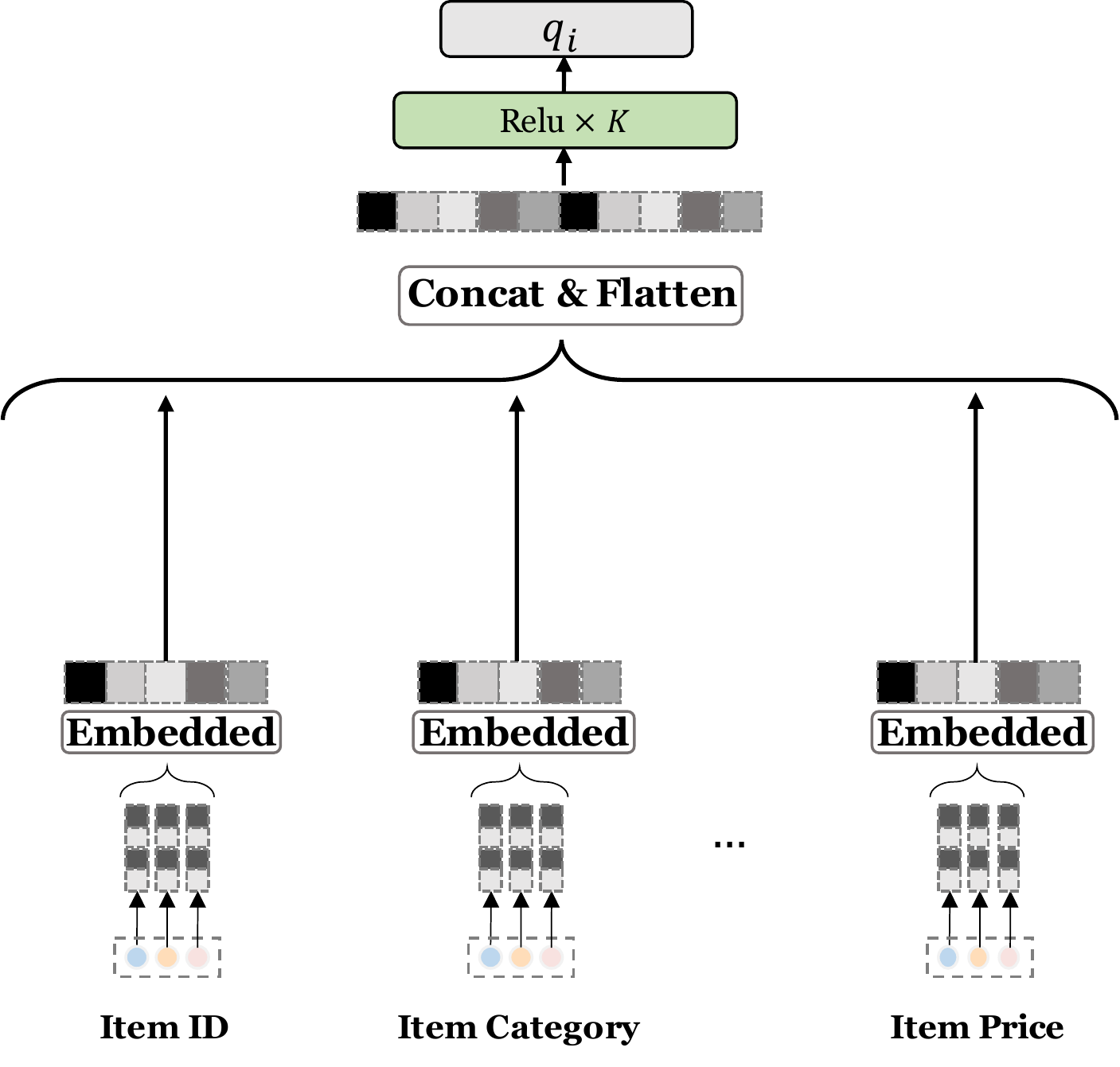}
        \label{fig:item_structure}
    }
    \caption{illustration of Causal collaborative filtering.}
    \label{fig:causal_cf}
\end{figure*}

Modern recommender systems have witnessed the success of the collaborative filtering---\textit{like-minded users will have similar interests to similar items}.
The two primary area of collaborative filtering are the neighborhood methods~\cite{user_cf1,user_cf2} and latent factor models~\cite{rec_mf}.
Neighborhood-based methods focus on computing the similarities between items or users.
Latent factor models alternatively try to explain the ratings by characterizing both items and users on latent factors inferred from
the rating patterns.
One successful realization of latent factor models is matrix factorization.
As shown in Figure~\ref{fig:mf_model}, by abstracting the user-item interactions as a rating matrix,
matrix factorization maps both users and items to a joint latent space of dimensionality $k$,
such that the interactions are predicted by the inner product in that space.
The simplest form is 
\begin{equation}\label{eq:mf}
    \hat{y}_{u,i} = \left(\mathcal{P} \mathcal{Q}^{\top}\right)_{u,i} = p_u^{\top} q_i
\end{equation}
where $p_u \in \mathbb{R}^k$ and $q_i \in \mathbb{R}^k$ are latent vectors associated with the user $u$ and item $i$.
With the learned representations of users and items, it can easily predict the rating a user will give to any item by equation~\eqref{eq:mf}.
In recent years, due to the prevalence of deep learning, 
the inner product can be replaced by the deep neural networks for the non-linearity interactions~\cite{ncf}.
And the representation of the user and item can be better encoded via sequential modeling~\cite{din,bert4rec} or graph neural network~\cite{LightGCN}.




\subsection{Causal Collaborative Filtering}\label{sec:causcf}

When it comes to the causal setting where the items can be purchased without recommendation, 
we interpret the assumption of causal collaborative filtering as \textit{like-minded users or similar items will behave similar causal effects of recommendation}.
Although each individual can not observe both the factual and counterfactual outcomes,
the causal effect estimation of users and items can benefit from the collaborative information from the similar users or items.
Inspired by matrix factorization, we materialize the latent factor model in causal collaborative filtering as the tensor factorization.
The rating tensor $R$ includes the dimensionality of the user, item, and treatment.
Since users can purchase items from the recommender system or other platform scenarios,
we have two rating channels corresponding to whether there is a recommendation or not.
Note that the rating value could be the binary value indicating whether to purchase or the continuous value of purchase quantity.
By this reformulation, 
the causal inference problem in recommendation becomes how to factorize the rating tensor and impute the missing values.
The typical factorization solutions include Tucker decomposition and CP decomposition,
but they are often computationally expensive or even infeasible due to the high sparsity of the decomposed data tensor~\cite{tensor_dec,context_rec1}.
Thus, we alternatively designed a much easier decomposition way via modeling the \textit{pairwise} interactions between such three dimensions and named it as the CausCF,
which is a natural extension of the matrix factorization.

Here, we describe our CausCF method.
Let $R=[y_{u,i,t}]$ be a 3-dimens\-ional rating tensor of shape $m \times n \times l$ with $m$ users, $n$ items, 
and $l$ different treatments.
Here, $l=2$ indicates the binary treatment setting while it is easy to extend for multiple treatments. e.g., multiple recommendation exposures.
Let $\mathcal{P} = [p_u]$ denote the user-factor matrix with the size of $m \times k$,
$\mathcal{Q} = [q_{i}]$ denote the item-factor matrix with the size of $n \times k$, 
and $\mathcal{D} = [d_{t}]$ be the treatment-factor matrix with the size of $l \times k$.
The notation $k$ is the rank of the latent factor model.
Figure~\ref{fig:infer_model} shows a simple example, where the rating tensor is of the shape $m=5, n=4, l=2$, and $k=3$.
Based on the principle of the pairwise interaction, the rating value $\hat{y}_{u,i,t}$ can be predicted as:
\begin{equation}\label{eq:pred}
    \begin{aligned}
    \hat{y}_{u,i,t} &=\left(\mathcal{P} \mathcal{Q}^{\top}\right)_{u,i}+\left(\mathcal{P} \mathcal{D}^{\top}\right)_{u,t}+\left(\mathcal{Q} \mathcal{D}^{\top}\right)_{i,t} \\ 
    &= p_u^\top q_i+ p_u^\top d_{t} + q_i^\top d_{t}
    \end{aligned}
\end{equation}
where $p_u \in \mathbb{R}^k, q_i \in \mathbb{R}^k $ and $d_t \in \mathbb{R}^k$ are latent factors of the user $u$, item $i$, and treatment $t$.
We implement these latent factors by embedding from one-hot representation or using feature-based non-linear projection functions
\begin{equation}
    p_u = f(x_u), \quad q_i = g(x_i), \quad d_t = h(t).
\end{equation}


In detail, we materialize the function $f(x_u)$ and $g(x_i)$ by the deep learning network and implement $h(t)$ by embedding from one-hot representation.
As shown in Figure~\ref{fig:user_structure}, $p_u$ is modeled by the user network with the inputs of the user attributes.
The attributes like age or gender are first embedded into the dense representations and then compressed by the fully connected layers.
The last layer's output with the dimensionality of $k$ is considered as the user latent representations.
Similarly, $q_i$ is extracted from the item network with the inputs of the item attributes, e.g., identifier, category, and price.
The attributes are also embedded into the dense representations and then fed into the fully connected layers.
The output vector is the final item representation.
Since the treatment is a single variable, we directly use its embedded representation, i.e., $h(t)=\mathrm{Embedded}(t)$, 
which has shown the effectiveness in the previous work when considering multiple treatments~\cite{multi_treatment}.
Note that any reasonable implementation of $p_u, q_i$ and $d_t$ can be plugged in our CausCF framework, e.g., the sequential model to enhance the user representation~\cite{sasrec,din}, and the graph embedding to enrich the item representation~\cite{LightGCN}.
How to choose a better implementation is not in the scope of this paper.

 \textbf{Discussion.}
Although the design of CausCF is much simpler than existing works using domain adaptation abstraction~\cite{tarnet,ace} or dedicated designed regularization terms~\cite{causal_emb}, 
it has competitive expressiveness.
If we connect CausCF to the contextual recommendation~\cite{context}, 
estimating the potential outcome under each treatment can be abstracted as predicting the rating value in different contexts.
So it is natural to formulate the problem as a tensor factorization task.
When considering it from the causal perspective, 
the prediction value consists of three components:
1) the rating value estimated by the user-item interaction $p_u^\top q_i$, which reflects the user preference towards the item;
2) the rating value estimated by the user-treatment interaction $p_u^\top d_t$, which describes the user-specific treatment effects;
3) the rating value estimated by the item-treatment interaction $q_i^\top d_t $, which explains the item-specific treatment effects.
Given the controlled user preference measured by $p_u^{\top} q_i$, 
similar users or similar items will behave similar treatment effects in our method.
If we compare the computational effort, CausCF is computationally more efficient.
We only model 2-dimensional interactions in an additive way without going through the high-order interactions as the existing tensor factorization solution~\cite{tensor_dec}.


\subsection{Training and Inference}

The decomposition in Equation~\eqref{eq:pred} is a natural extension of the matrix factorization in two dimensions.
So we can set up the optimization problem as the latent factor models.
Let $S$ denotes the set of all observed entries in $R$.
\begin{equation}
    S=\{(u, i, t): y_{u, i, t}\text{ is observed}\}.
\end{equation}
For each user-item pair, there is at most one observation under different treatments.
Since we center the problem on whether users make purchases on the platform, 
we denote the purchased entries by \textbf{1}s while the rest by \textbf{0}s.
The rating prediction problem is simplified to a binary classification task with the loss function of:
\begin{equation}
    \mathcal{L} = {-}\!\sum_{(u,i,t) \in S} \mkern-9mu\Bigl(y_{u,i,t}\!\log\! \left(\!\sigma\!\left(\hat{y}_{u,i,t}\right)\right){+}\left(1{-}y_{u,i,t}\!\right) \log\!\left(\!1{-}\sigma\!\left(\hat{y}_{u,i,t}\right)\right)\Bigr)
\end{equation}
where $\sigma$ is the sigmoid activation function to transform the predication value to the probability.

During the training phase, we optimize the representation of $p_u, q_i$, and $d_t$ by the stochastic gradient descent.
The $\ell_2$ regularization term is also added to the loss function to prevent overfitting.
In the prediction stage,
we try to infer all potential outcomes for each user-item pair by manually assigning the value of the treatments
\begin{equation}
    \begin{aligned}
        \hat{y}_{u, i, t=0} = &\, p_{u}^{\top} q_{i}+p_{u}^{\top} d_{t=0}+q_{i}^{\top} d_{t=0}, \\
        \hat{y}_{u, i, t=1} = &\, p_{u}^{\top} q_{i}+p_{u}^{\top} d_{t=1}+q_{i}^{\top} d_{t=1}.
    \end{aligned}
\end{equation}
The causal effect is then estimated by comparing the outcomes of different treatments
\begin{equation}
    \widehat{\mathrm{ITE}} = \hat{y}_{u, i, t=1} - \hat{y}_{u, i, t=0}.
\end{equation}

%% file: subfile/5_evaluation.tex
\section{Causal Effect Evaluation}\label{sec:discontinuity}

In this section, we will introduce a novel causal evaluation method via regression discontinuity design,
a typical econometric tool to make unbiased causal conclusions.

\subsection{Regression Discontinuity Design}

For causal effect evaluation,
researchers often rely on the dataset collected from the random policy or the simulated outcomes under different treatments~\cite{survey1}.
However, either conducting randomized experiments or simulating outcomes is impractical for a real-world recommender system.
Conducting randomized experiments is costly and even causes the risk of decreasing user engagements.
Simulation results are sometimes derived from the realistic setting.
A possible way is to design quasi-experiments to mimic the random policies.
In econometrics, the regression discontinuity design (RDD) is a typical method to analyze the causal effect in population, with the advantages of testable assumptions 
and tractable treatment assignments.
The main idea is that we could find a feature and a cutoff value to determine
or partially determine the status of the treatments~\cite{rd_2008,rd_2010}.
The average causal effect can be estimated by comparing observations lying closely on both sides of the cutoff value.
Formally, the treatment assignment follows
\begin{equation}
    T =\left\{\begin{array}{ll}
        1, & \text{ if } r <= c \\
        0, & \text{ if } r > c.
        \end{array}\right.
\end{equation}
where $r$ is the running variable deciding the assignment of treatments and $c$ is the cutoff value.
The causal effect $\mathrm{CATE}_c$ can be estimated in a non-parametric way:
\begin{equation}\label{eq:rdd}
    \begin{aligned}
        \mathrm{CATE}_c & {=} \mathbb{E}[Y(T=1) - Y(T=0)|r=c] \\
        &{=} \lim_{\epsilon \rightarrow 0} \mathbb{E}[Y(T{=}1)| r {=} c {-} \epsilon] {-} \lim_{\epsilon \rightarrow 0} \mathbb{E}[Y(T{=}0)| r {=} c {+} \epsilon] \\
        &{=} \lim_{\epsilon \rightarrow 0} \mathbb{E}[Y| r = c - \epsilon] - \lim_{\epsilon \rightarrow 0} \mathbb{E}[Y| r = c + \epsilon].
    \end{aligned}
\end{equation}

To explain the idea of the RDD, 
we first illustrate with the example of evaluating the effect of scholarship programs~\cite{rdd_first} and then make analogies for the recommender system.
To study the effect of scholarship programs on student performance,
a simple solution is to compare the outcomes of awardees and non-recipients.
But this comparison is problematic and may deduce an upward bias of the estimates.
Even if the scholarship did not improve grades at all, awardees would have performed better than non-recipients, 
simply because scholarships were given to students who had performed well before.
RDD exploits such exogenous characteristics of the intervention by constructing the following comparisons. 
If only students with grades above 90\% have scholarships,
it is possible to elicit the treatment effect by comparing students around the 90\% cutoff. 
The intuition here is that students scoring 89\% and 91\% are likely to be very similar, and the treatment effect is delivered by comparing the outcome of the students with 91\% (treated group) to the students with 89\% (control group).

\begin{figure}[t]
    \centering
    \includegraphics[width=.9\linewidth]{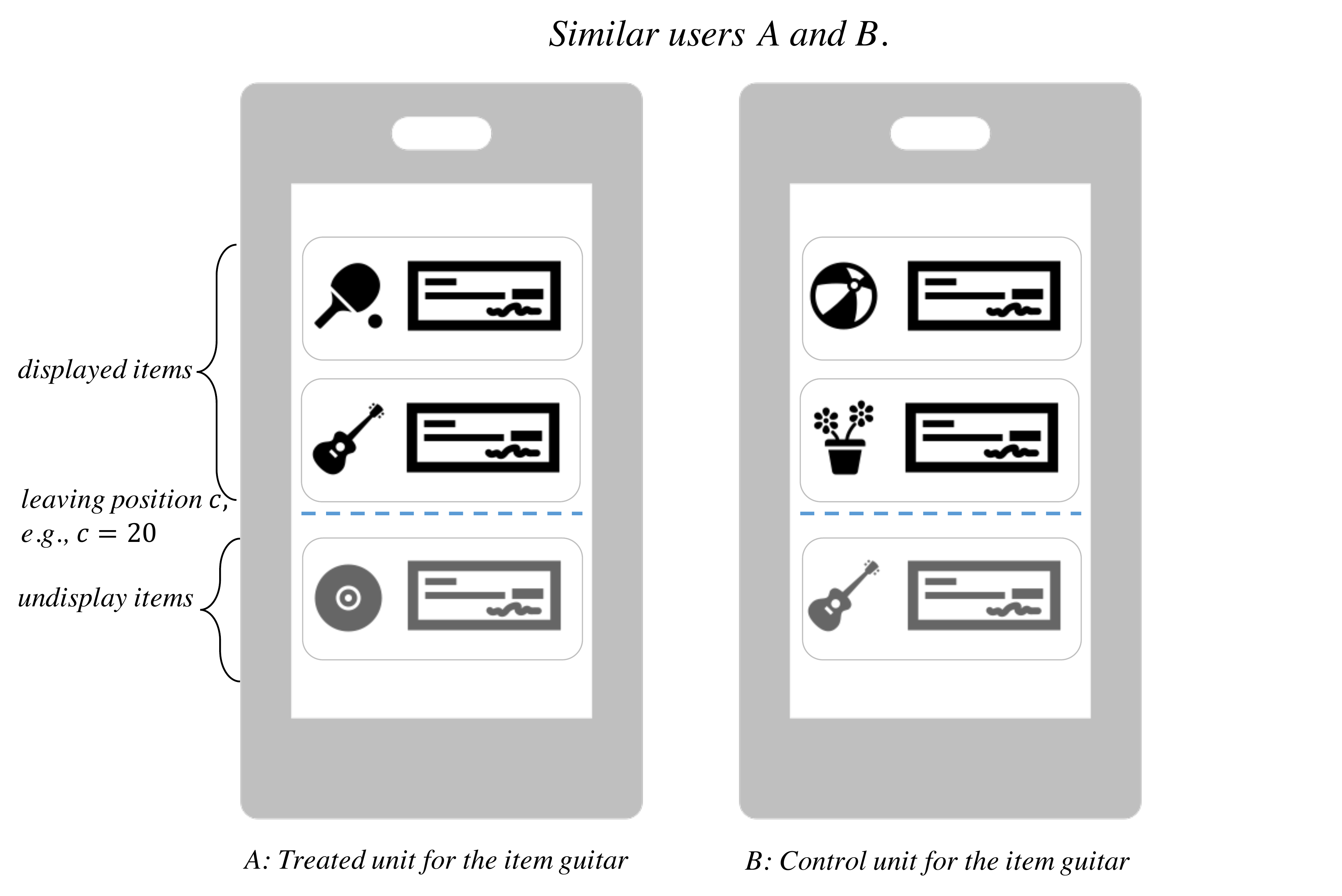}
    \caption{A typical recommender system with the streaming recommended items. 
    The items above the blue dotted line are successfully recommended while the rest are not.}
    \label{fig:rdd_case}
\end{figure}

When studying the causal effect of the recommendation, 
we have the following analogies.
As shown in Figure~\ref{fig:rdd_case},
given the item $i$, the item recommended $T=\{1, 0\}$ is determined by whether the users have browsed to the displayed position of the item.
If the item is displayed in the position before the user leaving the system, we consider it is successfully recommended to the user. Otherwise, it is not recommended.
Thus, the running variable $r$ can be defined as the displayed position of the item $i$ 
and the cutoff value $c$ is the position where the user stops browsing and leaves the system.
The recommendation effect of the item $i$ can be measured by comparing the observations around the leaving position.
The intuition here is that the users with the same item displayed in adjacent positions tend to be homogeneous.
Thus, the difference in the outcomes is mainly caused by whether the item is exposed or not.
Note that users can leave the system at any position, e.g., $c=20, 50,$ or $80$, 
so we will have multiple cutoff positions when studying an item's causal effect.
We conduct RDD analysis in each cutoff position and aggregate the causal conclusion by weighing the effect estimated at each position
\begin{equation}
    \mathrm{ATE} = \sum_c w_c \mathrm{CATE}_c.
\end{equation}
where $w_c$ is the percentage of users who leave the recommender system at the position $c$.

\subsection{Population Homogeneity Test} 


To apply RDD analysis in our experiments and use it for comparing the performance of different causal inference methods, 
we need to guarantee the validity of the RDD.
Generally, it relies on the assumption that those being barely treated are similar to those who are barely not treated.
For recommender systems, testing the similarity of populations is realized by the distribution test between user attributes.
Note that the user attributes are either discrete or continuous.
If we represent each discrete feature by the one-hot vector, 
it will be problematic to conduct the homogeneity test under such high dimensionalities.
Moreover, exiting hypothesis test methods mostly require that the input space follows a specific distribution,
e.g., normal distribution or log-normal distribution~\cite{ttest}.
When each dimension of the input space is heterogeneous and sampled from different distribution families,
it is hard to guarantee the confidence of the results~\cite{kernel_test}.
We therefore expect a low-dimensional well-defined space that can measure the homogeneity of the population effectively.
As illustrated in section~\ref{sec:causcf}, 
we use the compressed representation of users $p_u$ to conduct homogeneity test.
If the population distributions of $p_u$ around the cutoff position are balanced,
we think that the difference in the purchase probabilities is mainly caused by the item displayed or not.


\begin{figure}[t]
    \centering
    \includegraphics[width=.95\linewidth]{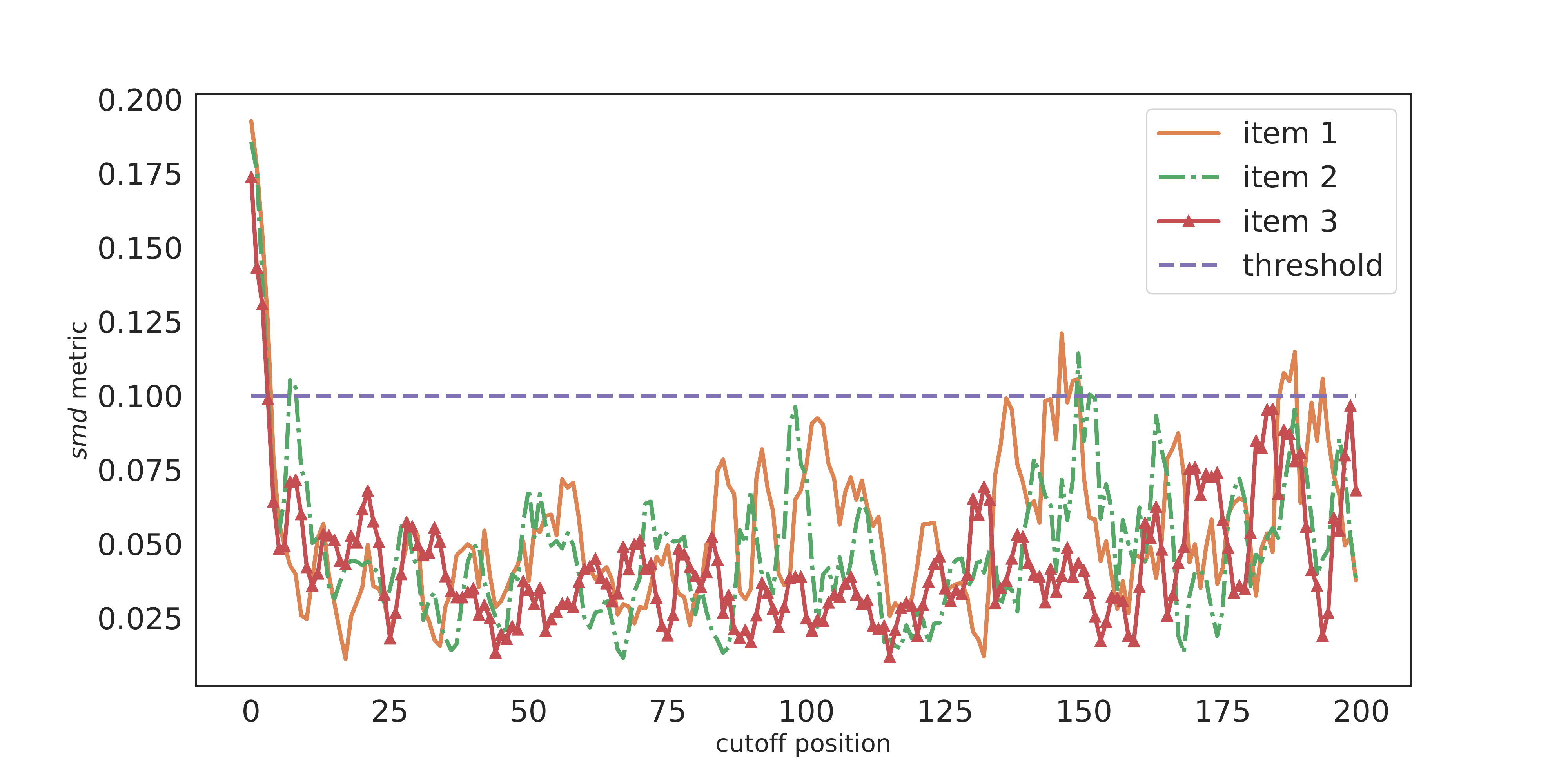}
    \caption{The population homogeneity test measured by the standardized difference metric. 
    The value that is less than 0.1 indicates the adequate balance of the distribution.
    }
    \label{fig:homogeneity_dist}
\end{figure}

As shown in Figure~\ref{fig:homogeneity_dist}, 
we randomly sampled three items from an e-commer\-cial platform (see more details about the dataset in section~\ref{sec:dataset}) and plot the results of the population homogeneity test w.r.t. each cutoff position.
In general, users browse 20-30 items and then leave the system. 
To avoid the bias of the conclusion caused by the sample size in long-tail positions, 
we only report the results of the top 200 positions for the homogeneity test.
We use the absolute standardized difference metric~\cite{smd_intro} to conduct the homogeneity test in each dimension of $p_u$ and then average the conclusion.
Formally, the absolute standardized difference is computed by:
\begin{equation}\label{eq:smd}
    \small
    \textit{smd}_j = \frac{ | \mu_{j,T=1} - \mu_{j,T=0} |}{\sqrt{0.5 \times (\sigma_{j, T=1}^2 + \sigma_{j, T=0}^2)}}, \quad \textit{smd} = \frac{1}{k} \sum_{j=1}^k \textit{smd}_j
\end{equation}
with the implications that the value less than 0.1 indicates the adequate distribution balance,
the values between 0.1 and 0.2 are not too alarming,
and the value greater than 0.2 means a serious imbalance of the distribution~\cite{smd}.
And the $\mu_{j,T=t}$ and $\sigma_{j, T=t}$ are the mean and standard deviation of the $j$-th dimension of $p_u$ in the corresponding treatment group.
The observations are summarized as:
1) as the user's browsing depth increases, the population around the cutoff points tends to be more homogeneous.
2) for the positions from 20 to 200, the test values are mostly under threshold 0.1,
which indicates the homogeneity of populations at these positions.
3) although there are higher variances in the last 50 positions, it is because of the limited sample size, according to our observations.

\begin{algorithm}[t]
    \small
\caption{RDD Analysis on Recommendation Effect}
\label{alg:rdd_iter}
\begin{algorithmic}[1] 
\State initialize the buffer $\mathcal{B}$ to save the causal effect for each item
\For{$i \sim \mathcal{I}$}
    \State initialize $w_i=[]$ and $v_i = []$
    \For{cutoff position $c \text{ in range} (s, e)$}
        \If {$\textit{smd}_c$ measured by Equation~\eqref{eq:smd} < 0.1}
            \State append the sample size around the cutoff position $c$ to $w_i$
            \State append $\mathrm{CATE}_c$ computed by Equation~\eqref{eq:rdd} to $v_i$
        \Else
            \State append $0$ to both $w_i$ and $v_i$
        \EndIf
    \EndFor
    \State append the item causal effect $\text{ATE} = \frac{\sum_c{w_i[c] * v_i[c]}}{\sum_c{w_i[c]}}$ to $\mathcal{B}$
\EndFor
\Return the buffer $\mathcal{B}$
\end{algorithmic}
\end{algorithm}

\subsection{Estimated Causal Effect}
Based on the result of the homogeneity test,
we can estimate the causal effect for different items with RDD analysis.
The computation process is summarized in Algorithm~\ref{alg:rdd_iter}.
The recommendation effect of each studied item can be measured by weighting the causal effect at each position.
Figure~\ref{fig:causal_cmp} plots the causal effect distribution of different items measured by RDD analysis.
As a comparison, we also plot the conclusions made by the data statistics,
which directly compares the purchase probability among the populations with the item recommended or not.
We observed that:
1) The distribution of the causal effect estimated by RDD analysis differs a lot from that by data statistics.
The causal effect estimated by RDD analysis is mostly less than 0.005, while the data statistics conclusions scatter from 0 to 0.01.
Figure~\ref{fig:causal_bias} presents the difference between the results obtained by RDD analysis and the data statistics.
Because the recommended items are always those that users will purchase with high probabilities,
the conclusion made by data statistics is often overestimated.
2) In Table~\ref{tab:top_ten_bias}, we list the top 10 items with the largest biases on recommendation effect.
The items with lower price or strong demand tend to arrive at a highly biased conclusion, e.g., 
Paper Diaper, Yoga Clothes, and Disposable Facecloth.
It is consistent with our common sense that users always have clear purchase intention for necessities, and they will purchase them even without recommendation.

%% file: subfile/6_experiment.tex
\section{Experiment}

In this section, we present how to build the dataset from an industrial recommender system for causal studies 
and then conduct experiments to answer the following research questions.

\begin{figure}[t]
    \centering
    \subfigure[Causal effect distribution comparison.]{
        \includegraphics[width=0.475\linewidth]{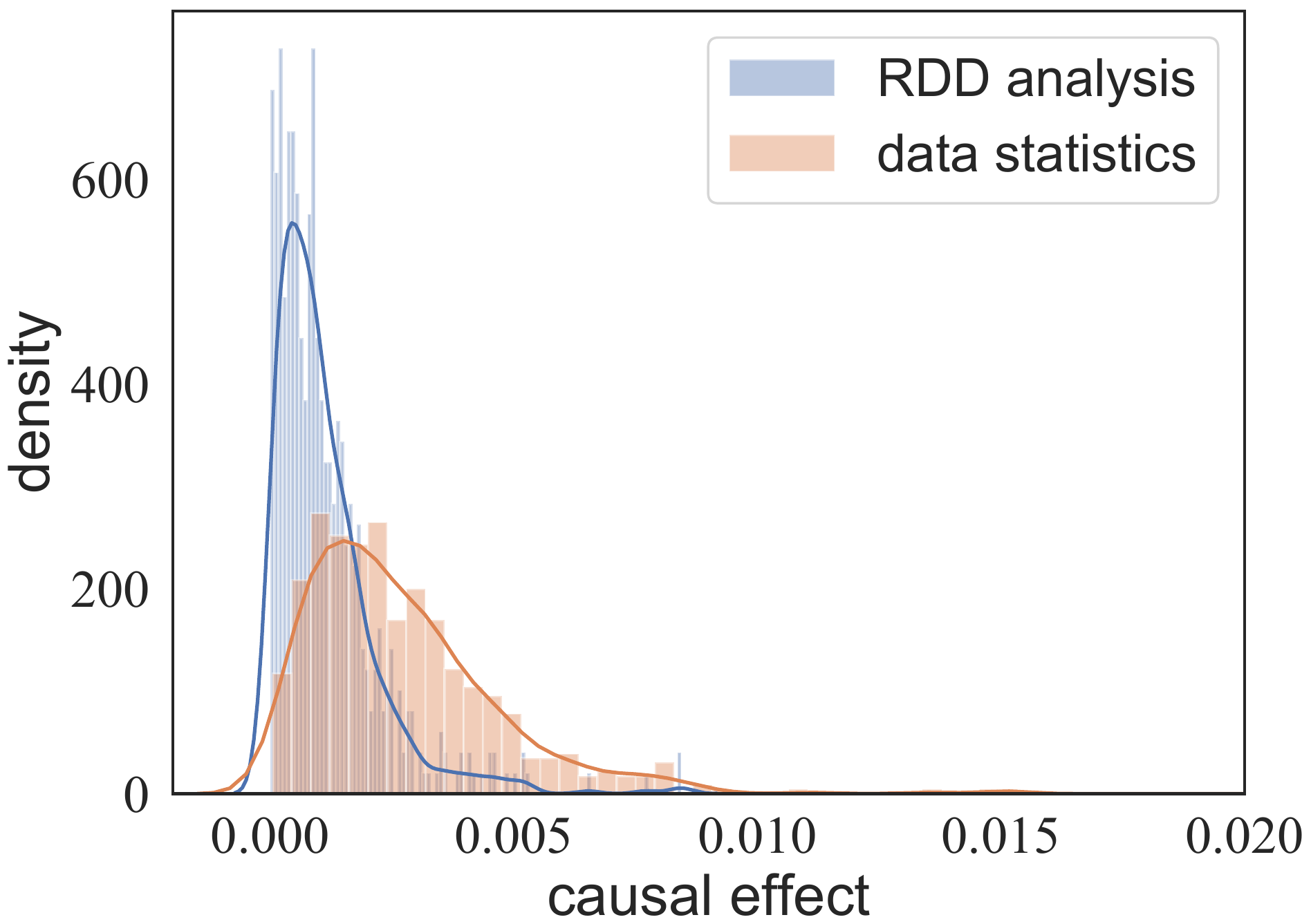}
        \label{fig:causal_cmp}
    }
    \subfigure[Statistic difference distribution.]{
        \includegraphics[width=0.475\linewidth]{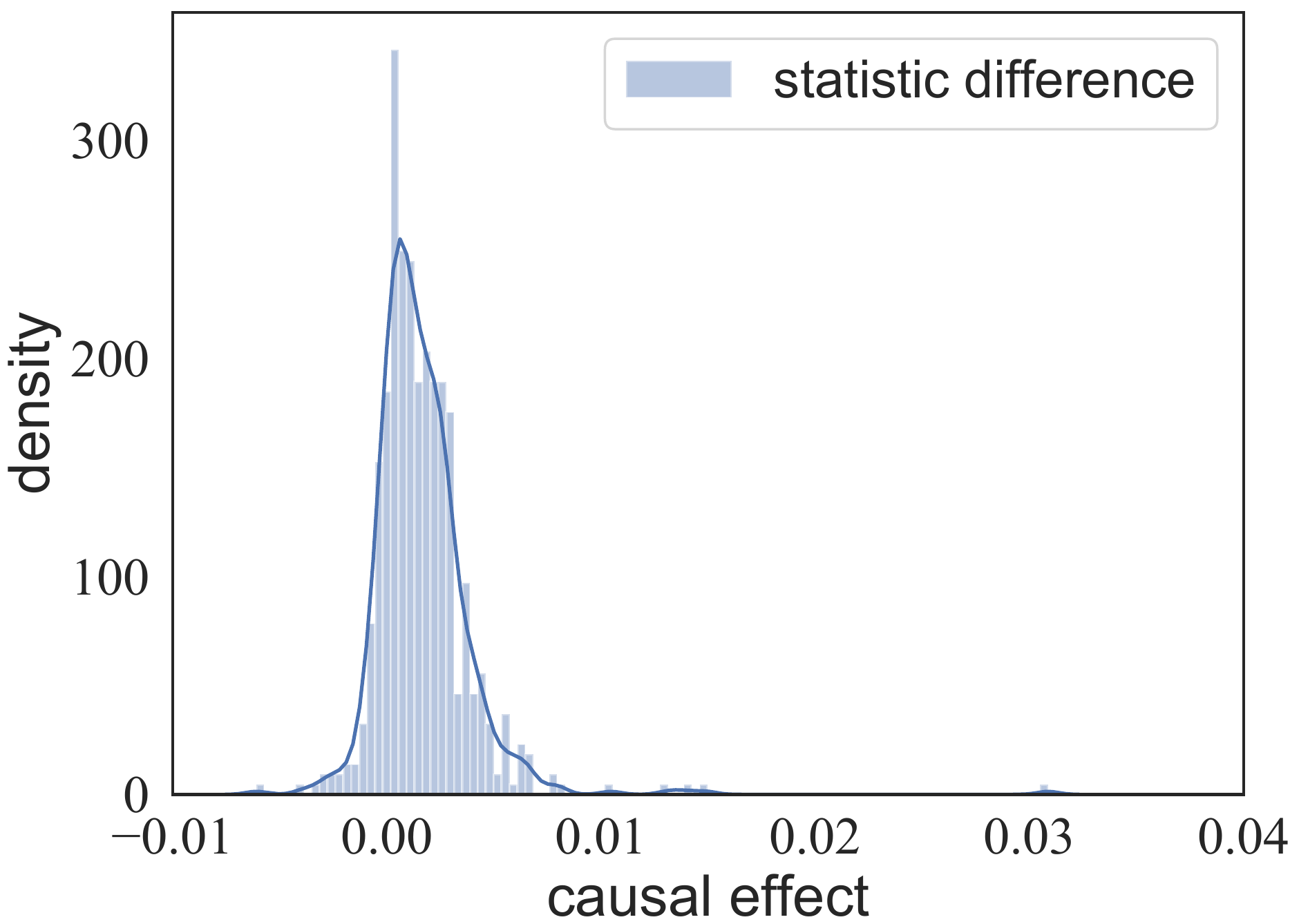}
        \label{fig:causal_bias}
    }
    \caption{Causal effect estimated by RDD analysis vs. conclusions made by data statistics.}
    \label{fig:causal_dist}
    \vspace{-10pt}
\end{figure}

\begin{table}[t]
    \centering
    \small
    \caption{Top ten items with the largest causal effects.}
    \label{tab:top_ten_bias}
    \begin{adjustbox}{max width=\linewidth}
    \begin{tabular}{l|ccc}
        \toprule
        Item & Data statistic & RDD analysis & Bias \\
        \midrule
        Paper Diaper & 0.06563 & 0.00389 & 0.06174 \\
        Instant Soup & 0.03872  & 0.00149 & 0.03723 \\
        Avocado & 0.03668 & 0.00226 & 0.03442 \\
        Sunscreen & 0.03101 & 0.00270 & 0.02830 \\
        Yoga Clothes & 0.03175 & 0.00398 & 0.02777 \\
        Cream Cleanser & 0.02856 & 0.00359 & 0.02497 \\
        Men Cleansing Gel & 0.02831 & 0.00482 & 0.02349 \\
        Baby Formula & 0.02344 & 0.00029 & 0.02315 \\
        Disposable Facecloth & 0.02527 & 0.00260 & 0.02267 \\
        Ginger Tea & 0.01932 & 0.00037 & 0.01894 \\ 
        \bottomrule
    \end{tabular}
    \end{adjustbox}
\end{table}

\begin{itemize}
\item \textbf{RQ1}. How does CausCF perform in terms of causal effect estimation comparing with SOTA causal inference methods?
\item \textbf{RQ2}. Does CausCF outperforms SOTA recommendation methods?
\item \textbf{RQ3}. How does CausCF perform for different populations?
\end{itemize}



\subsection{Experimental Settings}

\subsubsection{Dataset.}\label{sec:dataset}
We conduct experiments on an industrial dataset from Taobao and a publicly available dataset from Xing.

$\bullet$ \textbf{Taobao.}
The dataset is constructed from the event level logs of the interactions between users and platforms.
We focus on four types of data sources:
1) the recommendation logs from the ``Guess U Like" scenario in the homepages, including the displayed items, undisplayed items, and leaving positions.
It is worth mentioning that though there are also other recommendation channels in Taobao, we only focus on the causal effect of recommendations only in ``Guess U Like'' for simplicity.
Therefore, if the items are recommended from other channels, we regard them as ``non-recommended''
;
2) the user purchasing records from the whole platform with the details of transaction amount and quantity;
3) the user profile information and recent behaviors;
4) the item descriptions, e.g., title, category, and price.
The dataset is collected from 2020-07-10 to 2020-08-23, including 0.2 billion active users and 60 million items on the platform.
We merge these data sources according to the user and item identifiers.
In detail, the training and testing datasets for the causal study are constructed as follows.
We first consider each leaf category as the studied item because of the extraordinary data sparsity when dealing with the specific item on the platform---nearly 90\% items do not have any purchase records in the past one month. 
After this preprocessing, we have 2767 studied items in total.
Users are described by their demographic features and behaviors in the preceding ten days.
Items are described by their attributes and statistics.
The purchase amount or quantity is also simplified to the binary case.
Overall, we have 62 user-related features and 50 item-related features.
The training set consists of samples from 2020-08-10 to 2020-08-16 and is used to measure the within-sample performance of the estimated causal effect.
The test set is constructed from the following day 2020-08-17, and used for the out-of-sample evaluation.

$\bullet$ \textbf{Xing.} The dataset\footnote{\url{http://www.recsyschallenge.com/2017/}} contains user-job interactions at an online job-seeking site.
The interactions consist of impressions, clicks, bookmarks, applies, deletes, and recruiter interests.
We regard the positive interactions, clicks, bookmarks, and applies, as purchase behaviors and impressions as the platform recommendations.
The dataset is discretized by days and filtered according to the following conditions: items interacted by at least ten users and users logging at least ten days.
For interactions without recommendation, 
we randomly sampled unrecommended items for each user according to the ratio of 1 recommendation vs. 5 not recommendations.


\begin{table}[t]
    \small
    \centering
    \caption{Statistic of the dataset used in this paper.}
    \label{table:data_stat}
    \begin{adjustbox}{max width=\linewidth}
    \begin{tabular}{l r rrr}
        \toprule
        Dataset &  Users & Items & Purchases & Samples \\
        \midrule
        \multirow{2}{*}{Taobao}  & \multirow{2}{*}{232,949,123} & \multirow{2}{*}{2,767} & \multirow{2}{*}{764,550,731} & $1.03 \times 10^{10}$ treated \\
        & & & & $3.99 \times 10^{10}$ control \\
        \midrule
        \multirow{2}{*}{Xing} & \multirow{2}{*}{30,108} & \multirow{2}{*}{68,924} & \multirow{2}{*}{75,467} & $4.83 \times 10^6 $ treated\\
        & & & & $2.44 \times 10^7$ control \\
        \bottomrule
    \end{tabular}
    \end{adjustbox}
\end{table}

\subsubsection{Competitors for Causal Effect Estimation.}
To verify the effectiveness of the proposed method on causal effect estimation,
we compare it to the following competitors.
\begin{itemize}
    \item \textbf{Statistic}. The average treatment effect is measured by directly comparing the purchase probabilities between treated and control groups.
    \item \textbf{SNIPS}. The self-normalized estimator for counterfactual learning, 
    which measures the treatment effect by re-weighting the data samples with the estimated propensity score~\cite{snips}.
    \item \textbf{TARNet}. Treatment-agnostic representation network, 
    which separately predicts the potential outcomes by the treatment-private network~\cite{tarnet}.
    \item \textbf{CFR-MMD}. Counterfactual regression with MMD regularization~\cite{kernel_test} on distributions between the treated group and the control group.
    We add the MMD constrain on the output of the first fully connected layer~\cite{tarnet}.
    \item \textbf{CEVAE}. Causal effect variational auto-encoder, who introduces the latent variables to weaken the assumption on the data generating process and the unobserved confounders~\cite{cevae}.
\end{itemize}

\begin{table}[t]
    \centering
    \small
    \caption{Causal effect estimation comparison. The subscript with-s and out-of-s indicates the performance in the training set and testing set, respectively. }
    \label{table:result_cmp}
    \begin{adjustbox}{max width=\linewidth}
    \begin{tabular}{l|cc|cc}
        \toprule
        Method & $\mathrm{ATE}_{\text{within-s}}$ & $\mathrm{ATE}_{\text{out-of-s}}$ & $\epsilon_{\mathrm{ATE}}^{\text{within-s}}$ & $\epsilon_{\mathrm{ATE}}^{\text{out-of-s}}$ \\
        \midrule
        RDD  & 0.000741 & 0.000740 & -- & --  \\
        \midrule
        Statistic  & 0.002906 & 0.002900 & 0.002165 & 0.002160 \\
        SNIPS & 0.001735 & 0.001714 & 0.000994 & 0.000974 \\
        TARNet  & 0.001833 & 0.001828 & 0.001092 & 0.001088 \\
        CFR-MMD & 0.001639 & 0.001634 & 0.000898 & 0.000894 \\
        CEVAE  & 0.002279 & 0.002275 & 0.001538 & 0.001535 \\
        CausCF  & \textbf{0.001515} &  \textbf{0.001481} & \textbf{0.000774} & \textbf{0.000741} \\
        \bottomrule
    \end{tabular}
    \end{adjustbox}
\end{table}

\subsubsection{Competitors for Ranking Performance. }
We also compare our method with other ranking competitors to observe the differences of the recommender systems that rank the items according to the predicted purchase probabilities or causal effects.
\begin{itemize}
    \item \textbf{NCF}. The matrix factorization implementation where the inner product function is replaced by deep neural network~\cite{ncf}.
    \item \textbf{WDL (Wide\&Deep)}. It models the low-order and high-order feature interactions simultaneously. 
    The wide side is a linear regression, and the deep side is a neural network~\cite{WDL}.
    \item \textbf{CausE}. We jointly train two NCFs with and without recommendation. 
    The causal effect is measured by the difference between the predicted value of two NCFs~\cite{causal_emb}.
    \item \textbf{CausE$_i$}. The variant of CausE with shared user factors and private item factors for treated group and control group~\cite{causal_emb}.
    \item \textbf{ULRMF}. The NCF trained with the uplift-based sampling strategy~\cite{rec_uplift}.
\end{itemize}

\subsubsection{Evaluation Metrics.}

Since we only have the observation that is either treated or controlled,
we measure the performance by the absolute error of the ATE in the population.
Here, we consider the ATE metric measured by RDD analysis as the ground truth.
In section~\ref{sec:discontinuity}, 
we have analyzed the unbiasedness of RDD analysis when the randomized experiment is infeasible.
Formally, the absolute error of the ATE is calculated as follows:
\begin{equation}
    \epsilon_{\mathrm{ATE}} = \Bigl|\mathrm{ATE}_{\mathrm{RDD}} - \frac{1}{m \times n} \sum_{(u, i)}(\hat{y}_{u, i, t=1} - \hat{y}_{u, i, t=0})\Bigr|.
\end{equation}

Except for the precision of the estimated causal effect, 
we also concern about the ranking performance when sorting the list according to the causal effect.
Here, we refer to the uplift metric proposed by ~\cite{rec_uplift} to evaluate the ranking performance.
\citet{rec_uplift} argued that the evaluation of the recommender system should not focus on the accuracy-based metric like precision (\# of purchased items vs. \# of recommended items)
but the uplift of the purchasing willingness by recommendation.
Formally, the uplift metric is computed by
\begin{equation}
    \small
    \begin{aligned}
    \hat{\tau}_{L_{u}^{M}} &=\frac{1}{\left|L_{u}^{\scaleto{M \cap D}{2.8pt}}\right|} \sum_{i \in L_{u}^{\scaleto{M \cap D}{2.8pt}}}  y_{u,i,t=1} - \frac{1}{\left|L_{u}^{\scaleto{M \backslash D}{3.6pt}}\right|} \sum_{i \in L_{u}^{\scaleto{M \backslash D}{3.6pt}}} y_{u,i,t=0} \\
    \mathrm{Uplift} &\equiv \frac{1}{|U|} \sum_{u \in U} \hat{\tau}_{L_{u}^{M}}
    \end{aligned}
\end{equation}
where $L_u^M$ is the top $N$ ranking items of the recommendation method $M$ and $L_u^D$ is the existing recommendation loggings of the collected data $D$.
Thus, $L_u^{\scaleto{M \cap D}{2.8pt}}$ is the item set that has been successfully recommended while $L_u^{\scaleto{M \backslash D}{3.6pt}}$ is the set of the control group without recommendation.
The difference of the outcomes in $L_u^{\scaleto{M \cap D}{2.8pt}}$ and $L_u^{\scaleto{M \backslash D}{3.6pt}}$ describes the purchase lift of methods.
$\mathrm{Uplift}_{\mathrm{SNIPS}}$ is the variant integrating the propensity score (see more details in~\cite{rec_uplift}).

\begin{table}
    \centering
    \caption{Ranking performance comparison (the larger value is with the better performance).}
    \subtable[Taobao Dataset.]{
    \begin{adjustbox}{max width=\linewidth}
        \begin{tabular}{lccc ccc c}  
            \toprule
            \multirow{2}{*}{Model} & \multicolumn{3}{c}{$\text{Uplift}$} & \multicolumn{3}{c}{$\text{Uplift}_{\text{SNIPS}}$} & Precision \\
            \cmidrule(lr){2-4} \cmidrule(lr){5-7} \cmidrule(lr){8-8}
            & $N=10$ & $N=30$ & $N=50$ & $N=10$ & $N=30$ & $N=50$ & $N=10$ \\
            \midrule
            NCF & 0.01105 & 0.01115 & 0.01130 & 0.00779 & 0.00627 & 0.00587 & \textbf{0.02715} \\
            WDL & 0.01002 & 0.01086 & 0.01120 & 0.00708 & 0.00614 & 0.00583 & 0.02713 \\
            \midrule
            CausE & 0.01497 &  0.01169 & 0.01112 & 0.01028 & 0.00651 & 0.00579 & 0.01965 \\
            CausE$_i$ & 0.01641 &  0.01341 & 0.01242 & 0.01121 & 0.00732 & 0.00632 & 0.01653 \\
            ULRMF  & 0.00826 &  0.01213 & 0.01307 & 0.00578 & 0.00687 & 0.00672 & 0.02619 \\
            \midrule
            CausCF$_{u}$ & 0.01099 & 0.01075 & 0.01086 & 0.00766 & 0.00601 & 0.00563 & 0.02436 \\
            CausCF$_{i}$ & 0.01698 & 0.01437 & 0.01351 & 0.01149 & 0.00766 & 0.00668 & 0.02121 \\
            CausCF & \textbf{0.01816} & \textbf{0.01548} & \textbf{0.01435} & \textbf{0.01215} & \textbf{0.00828} & \textbf{0.00712} & 0.02054 \\
            \bottomrule
        \end{tabular}
        \end{adjustbox}
    }
    
    \subtable[Xing Dataset.]{
        \begin{adjustbox}{max width=\linewidth}
        \begin{tabular}{lccc ccc c}  
            \toprule
            \multirow{2}{*}{Model} & \multicolumn{3}{c}{$\text{Uplift}$} & \multicolumn{3}{c}{$\text{Uplift}_{\text{SNIPS}}$} & Precision \\
            \cmidrule(lr){2-4} \cmidrule(lr){5-7} \cmidrule(lr){8-8}
            & $N=10$ & $N=30$ & $N=50$ & $N=10$ & $N=30$ & $N=50$ & $N=10$ \\
            \midrule
            NCF & 0.00139 & 0.00209 & 0.00209 & 0.00140 & 0.00213 & 0.00212 & \textbf{0.00491}  \\
            WDL & 0.00142 & 0.00205 & 0.00210 & 0.00144 & 0.00209 & 0.00213 & 0.00491 \\
            \midrule
            CausE & 0.00227 & 0.00232 & 0.00215 & 0.00227 & 0.00235 & 0.00218 & 0.00425 \\
            CausE$_i$ & 0.00219 & 0.00219 & 0.00213 & 0.00225 & 0.00223 & 0.00216 & 0.00426 \\
            ULRMF  & 0.00218 & 0.00234 & 0.00214 & 0.00217 & 0.00235 & 0.00216 & 0.00468 \\
            \midrule
            CausCF$_{u}$ & 0.00189 & 0.00216 & 0.00212 & 0.00191 & 0.00220 & 0.00215 & 0.00477 \\
            CausCF$_{i}$ & 0.00270 & \textbf{0.00241} & 0.00214 & 0.00269 & \textbf{0.00241} & 0.00217 & 0.00430 \\
            CausCF & \textbf{0.00277} & 0.00240 & \textbf{0.00216} & \textbf{0.00275} & 0.00240 & \textbf{0.00218} & 0.00441  \\
            \bottomrule
        \end{tabular}
        \end{adjustbox}
    } 
    \label{tab:rank_cmp}
\end{table}

\begin{figure}[t]
    \centering
    \subfigure[Male advantage items w.r.t. the estimated causal effect.]{
        \includegraphics[width=0.42\linewidth,height=0.35\linewidth]{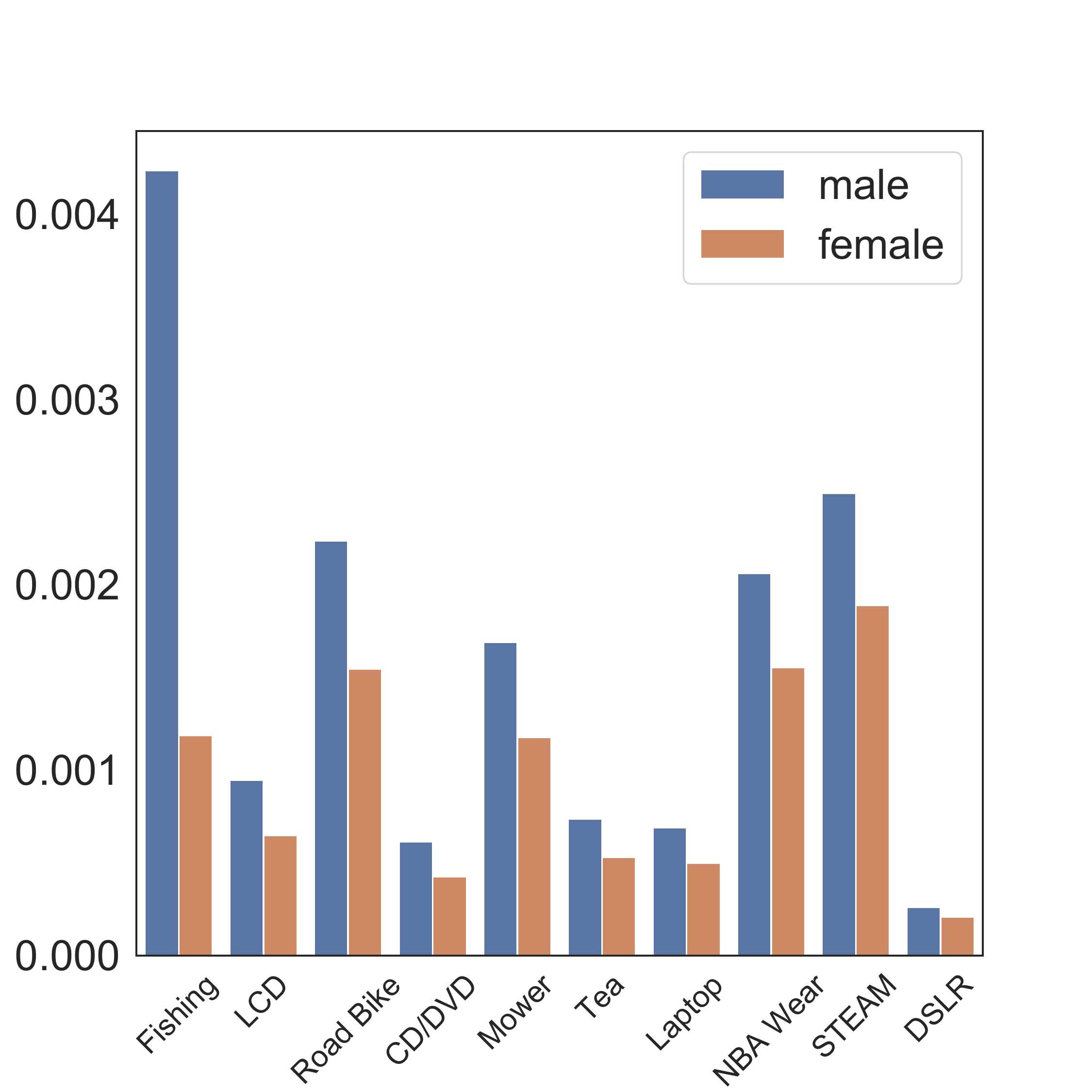}
        \label{fig:gender_male}
    }
    \hfill
    \subfigure[Female advantage items w.r.t. the estimated causal effect.]{
        \includegraphics[width=0.42\linewidth,height=0.35\linewidth]{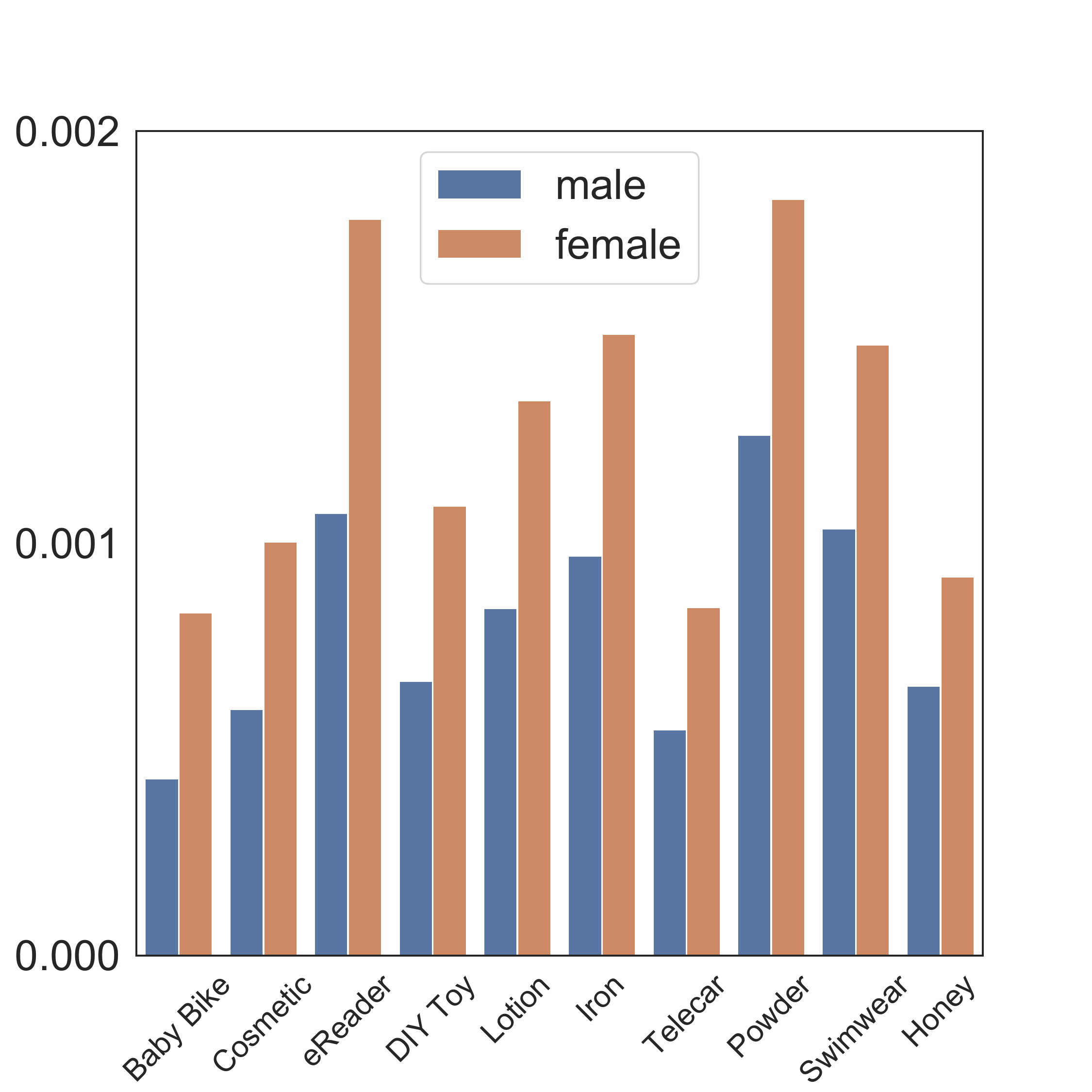}
        \label{fig:gender_female}
    }
    
    \subfigure[Age group 18$\sim$22 advantage items w.r.t. the estimated causal effect.]{
        \includegraphics[width=0.42\linewidth,height=0.35\linewidth]{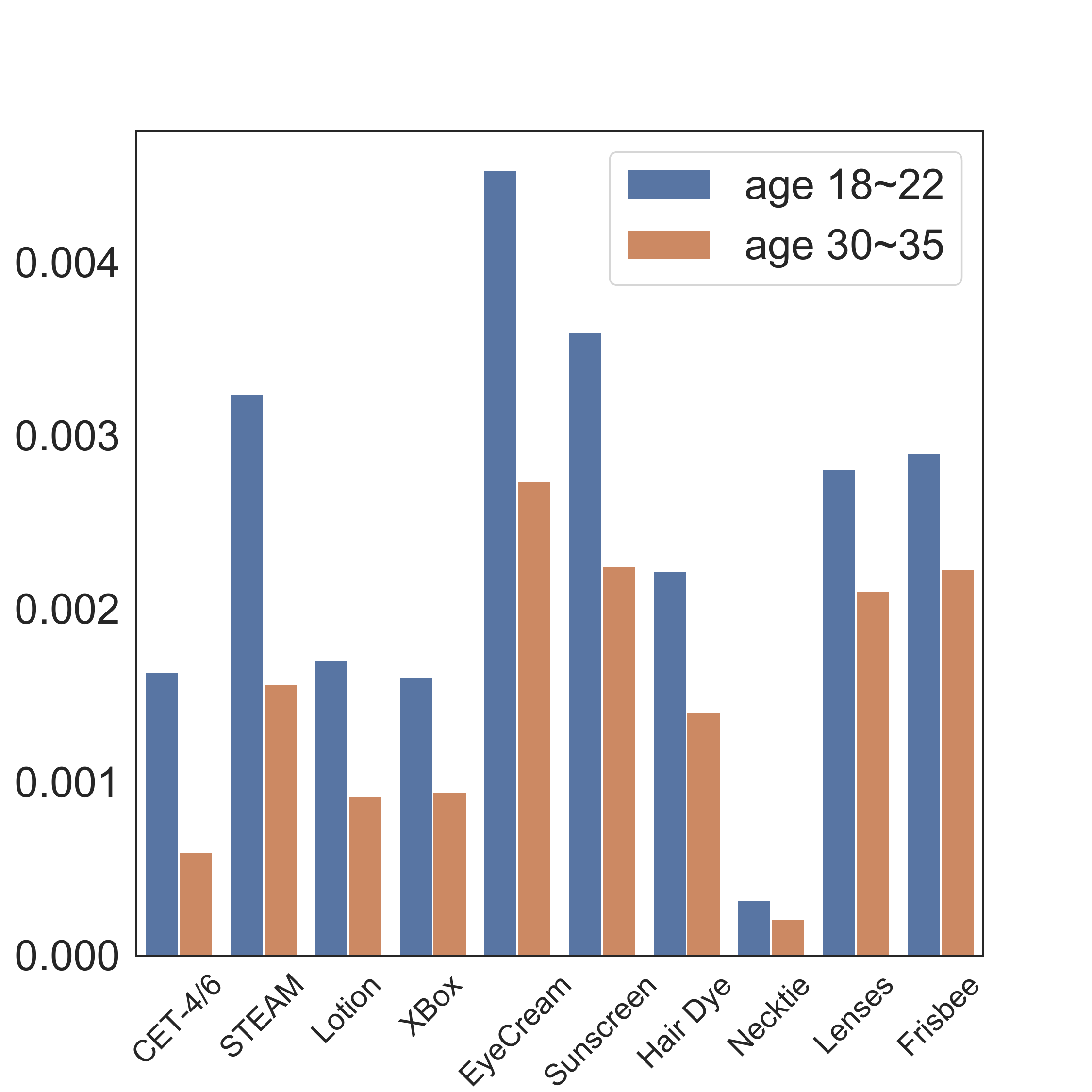}
        \label{fig:age_18}
    }
    \hfill
    \subfigure[Age group 30$\sim$35 advantage items w.r.t. the estimated causal effect.]{
        \includegraphics[width=0.42\linewidth,height=0.35\linewidth]{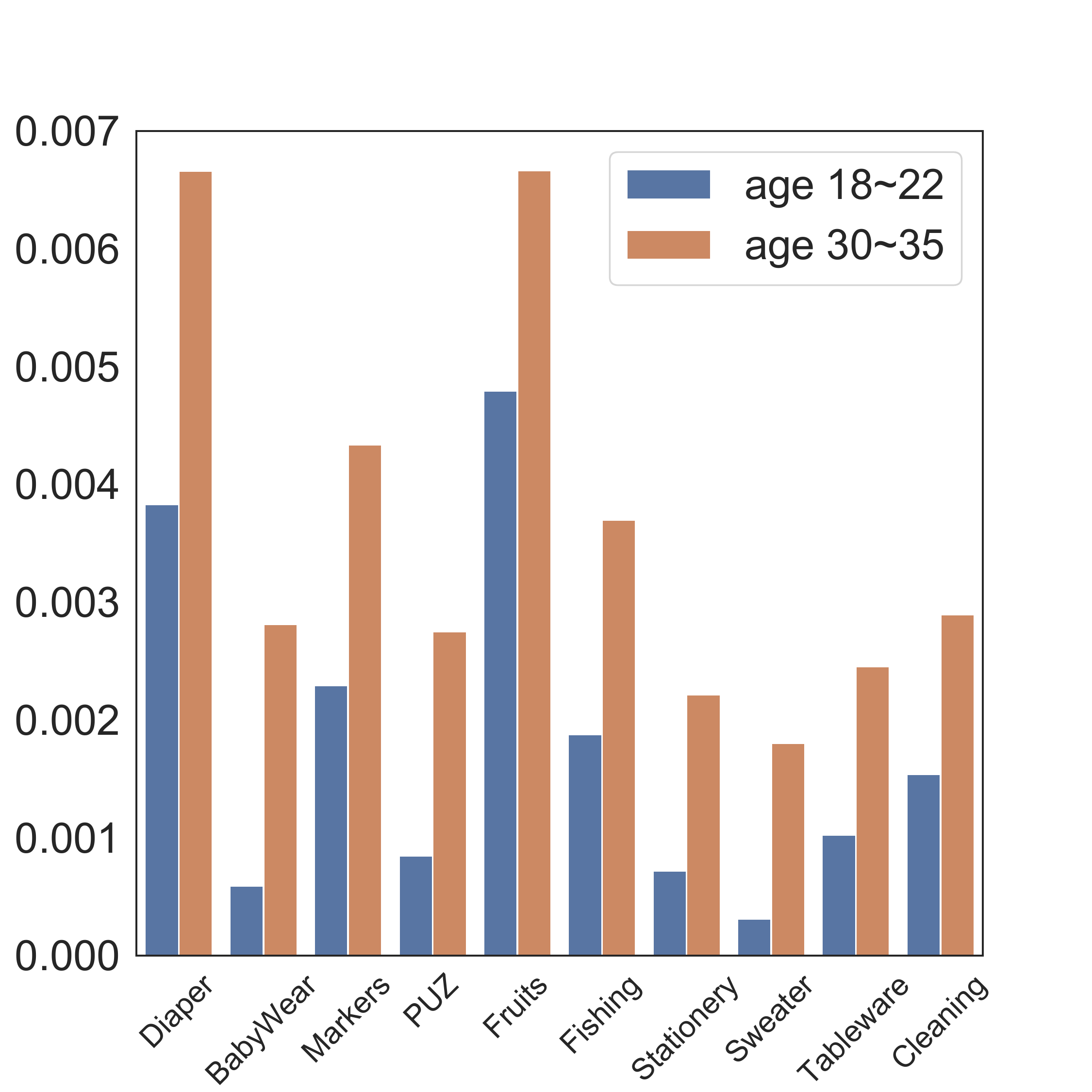}
        \label{fig:age_30}
    }
    \caption{Causal effect comparison among different groups.}
    \vspace{-8pt}
    \label{fig:gender_cmp}
\end{figure}

\subsubsection{Implementation Details.}
We conduct all experiments based on the implements using the Tensorflow library.
For TARNet and CFR-MMD, the publicly code\footnote{https://github.com/clinicalml/cfrnet} is available.
For CEVAE, we refer to the implementation from the github\footnote{https://github.com/AMLab-Amsterdam/CEVAE}.
To deal with the large data scale of the Taobao dataset, we update these implementations in a distributed training platform with 10 parameter servers and 201 training workers~\cite{abadi2016tensorflow}.
The batch size is set to 512 and 128 for Taobao and Xing dataset.
We adopt the Adagrad~\cite{adagrad} optimizer and set the initial learning rate as 0.001.
By default, the embedding dimension of each feature is 8.
The number of units in hidden layers is configured as $512 \rightarrow 256 \rightarrow 128$.

\subsection{Causal Effect Comparision (RQ1)}

Since the Xing dataset does not provide user browsing position information, 
we evaluate the precision of causal effect estimation mainly in Taobao dataset.
As shown in Table~\ref{table:result_cmp}, the first row is the conclusion made by the RDD, as the ground truth to compare different methods.
Except for the absolute error on the ATE metric, 
we supply the estimated ATE value in both the training set ($\mathrm{ATE}_{\text{within-s}}$) and the testing set ($\mathrm{ATE}_{\text{out-of-s}}$).
We find the recommendation effect in a commercial platform is around 0.00074 by our RDD analysis.
In other words, thousands of recommendations can lead users to a deal on the platform.

By comparing different methods, we have the following observations.
1) All compared methods have achieved promising generalization performance. There is no significant difference in the evaluation metric in the training set and testing set.
2) The data statistics arrived at the most biased conclusion, which is four times larger than that of RDD analysis,
because the user preference, as the unignorable confounder in the estimation, not only affects the recommendation policy of the platform, 
but the purchase willingness of users.
3) The propensity score is an empirical method to eliminate biases in previous works~\cite{rec_causal_icml,double_miss}. 
We find it is also plausible to measure the causal effect of the recommendation.
3) TARNet and CFR-MMD are competitive methods to make an unbiased conclusion. 
With the extra distribution regularization between the treated group and the control group, 
both the within-sample and out-of-sample performance improves.
4) CEVAE results are not satisfactory as we expect.
We consider that it is an extremely hard task to mimic the data generation process under high-dimensional spaces.
Most model efforts focus on the reconstruction of the pre-treatment variables
rather than the prediction of potential outcomes.
5) Our proposed causal collaborative filtering method achieved the best performance in the causal effect estimation task.
With the controlled user preferences,
we explicitly model the user and item-specific treatment effect,
which is more computationally efficient and intuitive to explain the user-item interactions. \vspace{-10pt}


\subsection{Ranking Performance Comparision (RQ2)}

Besides, the precision of the estimated causal effect, 
we also consider the impact on online ranking performance if ranking the items by causal effects.
Table~\ref{tab:rank_cmp} summarized the ranking performance of different recommender methods.
Our proposed CausCF method has achieved the best performance in both the public dataset and industrial application.
Compared with NCF and WDL methods aiming to optimize the accuracy-based metric, e.g., precision,
our method achieved significant improvements on the uplift metric.
A similar improvement can be observed on CausE and ULRMP methods,
since they are also designed for causal effect optimization.
But jointly training two NCFs for the treated group and the control can not effectively model the user and item-specific treatment 
and fail to utilize the observations from different treatment groups.
ULRMF is a method to optimize the causal effect directly while tuning the sampling ratio for not recommended and not purchased observations is a computationally expensive process.
When comparing the variants of CausCF that only model the user-specific treatment effect (CausCF$_u$) or the item-specific treatment effect (CauCF$_i$),
we find that the treatment effect is more sensitive for the item side.
Some previous works also proved the bias from items, e.g., popularity bias, is one of the main biases in recommender systems~\cite{pop_bias}.

\subsection{Case Studies on Populations (RQ3)}

We also explore the characteristics of different population groups when ranking the items by the causal effect.
The subgroup causal effect is computed by averaging the estimated ITE in that group.
Figure~\ref{fig:gender_male} and \ref{fig:gender_female} plot the top 10 advantage items for males and females with the largest causal effect difference.
It is interesting to see that the male's advantage items include fishing, road bike, tea, steam game, etc.
These items are mostly related to interests and hobbies.
More recommendations will lead to higher purchase probabilities.
The advantage items for the female are more related to the children and the beauty.
In fact, Chinese women recently pay more and more attention to skincare, beauty, and children's education.
Similarly, Figure~\ref{fig:age_18} and \ref{fig:age_30} plot the comparisons between different ages.
The users aged 18 to 22 are mostly college students, who are more interested in review materials for CET-4/6 test and fashion items.
In contrast, a large proportion of the users aged 30 to 35 are new parents, who are more likely to purchase the goods for children or the necessities under the recommendation.



%% file: subfile/7_conclusion.tex
\section{Conclusion}

In this paper, we proposed a novel causal inference method delicately designed for the recommender system.
By formulating the problem as a tensor factorization task, 
we realized modeling the user and item-specific treatment effect simultaneously. 
We also propose an evaluation approach--RDD analysis, which is practical when conducting randomized experiments is costly or even infeasible.
With the advantages of testable assumptions, the unbiasedness of the conclusion made by RDD analysis can be guaranteed.
The experiments on both the public dataset and industrial application have confirmed that our method outperforms existing causal inference methods and conventional accuracy-based recommendation algorithms.
The case studies on different groups of populations also shed light on the potential of the recommender system that optimizes causal effects.